\documentclass[10pt, one column, draft cls]{ieeeconf}
\IEEEoverridecommandlockouts 

\usepackage{amsmath,amssymb,bbm,color,psfrag,amsfonts,mathtools}
\usepackage{dsfont}
\usepackage{graphicx,cuted}                       
\usepackage[pdftex,pdfpagelabels,bookmarks,hyperindex,hyperfigures]{hyperref}
\usepackage{wrapfig}
\usepackage{pdfpages}
\usepackage[ruled,lined]{algorithm2e} 
\usepackage{algpseudocode}
   \SetKwInOut{Input}{input}\SetKwInOut{Output}{output}
\usepackage{accents}
\usepackage{lipsum}
\usepackage[T1]{fontenc}
\usepackage[utf8]{inputenc}
\usepackage{mathtools}
\usepackage{multicol}
\usepackage{paralist}   
\usepackage{kbordermatrix}

\usepackage{booktabs}  
\usepackage{multirow}
\usepackage[normalem]{ulem}

\usepackage{caption}
\usepackage{subcaption}

\usepackage{cite}

\newcounter{problem}

\newtheorem{theorem}{Theorem}

\newtheorem{lemma}{Lemma}
\newtheorem{proposition}{Proposition}

\newtheorem{assumption}{Assumption}

\newtheorem{remark}{Remark}

\usepackage{color}

\newcommand{\real}{\mathbb{R}}

\newcommand{\RR}{\mathbb{R}}

\newcommand{\mc}{\mathcal}

\newcommand{\ksmarginphantom}[1]{}

\newcommand{\setdef}[2]{\left\{#1 \; | \; #2\right\}}

\newcommand{\prb}{\mu_0}

\newcommand{\congfunc}{\ell}
\newcommand{\diag}{\mathbf{diag}}

\newcommand{\prior}{\mu_0}
\newcommand{\state}{\omega}
\newcommand{\allstates}{\Omega}
\newcommand{\interior}{\mathrm{int}}
\newcommand{\signal}{\pi}

\newcommand{\pfrac}{\nu}
\newcommand{\simplex}{\mc P}

\newcommand{\nlinks}{n}
\newcommand{\npaths}{\nlinks}

\newcommand{\nstates}{s}

\newcommand{\de}{\mathrm{d}}

\graphicspath{{../results/} , {../../../../fig/},{../fig/}}

\algnewcommand{\algorithmicgoto}{\textbf{go to}}%
\algnewcommand{\Goto}[1]{\algorithmicgoto~\ref{#1}}%

\usepackage{tikz}
\usetikzlibrary{positioning}
\usetikzlibrary{arrows}
\usepackage{verbatim}
\usepackage{soul}

\title{Convergence in a Repeated Non-atomic Routing Game \\ with Partial Signaling}

\author{Yixian Zhu\thanks{The authors are with the University of Southern California, Los Angeles, CA. \texttt{\{yixian,ksavla\}@usc.edu}. This work was supported in part by NSF CAREER ECCS \# 1454729. K. Savla has financial interest in Xtelligent, Inc.} \qquad Ketan Savla}

\date{\today}

\begin{document}
\maketitle

\begin{abstract}
We study the following repeated non-atomic routing game. In every round, nature chooses a state in an i.i.d. manner according to a publicly known distribution, which influences link latency functions. The system planner makes private route recommendations to participating agents, which constitute a fixed fraction, according to a publicly known signaling strategy. The participating agents choose between obeying or not obeying the recommendation according to cumulative regret of the participating agent population in the previous round. The non-participating agents choose route according to myopic best response to a calibrated forecast of the routing decisions of the participating agents. We show that, for parallel networks, if the planner's signal strategy satisfies the obedience condition, then, almost surely, the link flows are asymptotically consistent with the Bayes correlated equilibrium induced by the signaling strategy.  
\end{abstract}

\section{Introduction}
Optimal decision-making for agents in modern complex socio-technical systems, such as transportation, is challenging because of dynamical environments and uncertainties. Personal mobile devices
allow the agents to stay updated on the information, and even make decisions for them. This architecture also allows the possibility for \emph{information design}, whereby the source of information, say a system planner, can strategically pick what information to reveal or make direct route recommendations 
to steer the traffic flow towards optimal social state.

\ksmarginphantom{need to pick one among agent/player/driver and stick to it}
We consider a non-atomic routing game where the link latency functions are conditional on the state of the nature, e.g., whether there is a traffic incident or not. This state is generated in an i.i.d. manner from an exogenous distribution. This distribution is known publicly to all the agents, e.g., from historical data or personal experience, but the actual realization is known only to the system planner. It is therefore natural to consider that the system planner could use this informational advantage to recommend routes to the agents in order to optimize a social objective. Such a mapping from the state realization to route recommendation is referred to as signaling strategy in line with the \emph{information design} literature~\cite{Bergemann.Morris:19}. 
The recommendations would be obeyed by the agents only if, by following it, they are better off than by making decision based only on the prior distribution of the states. Such a requirement on the signaling strategy is formally referred to as \emph{obedience constraint}. 
Extension of this constraint to a multi-round setting is not clear. Specifically, how does an agent determine obedience based on the series of recommendations and payoffs that she receives, and can a static notion guarantee persistent obedience in a multi-round setting? One naive justification is that the agents follow the recommendations in every round until they have sufficient samples to empirically evaluate the obedience condition. If the strategy indeed satisfies the obedience constraint, then the agents' evaluation would be so and therefore they will continue following the recommendation subsequently. We rather consider a setting where the agents choose between obeying or not obeying recommendation in every round from the beginning. This choice is determined by the agents' regret associated with the decision in the previous round and aggregated over all other agents. 
Moreover, we also allow for partial signaling, i.e., when only a fraction of agents participate in signaling. The rest are assumed to choose routes as best response to their forecast of the link flows induced by the participating agents.  

The motivation for the setting comes from the practical consideration that agents have the discretion to use or not to use a navigation system, and if they use one, then whether to obey the recommendation or not on any given trip. It is therefore of interest to understand the long term effectiveness of a route recommendation strategy in such a setting. The feature of our model to let the decision of a participating agent depend on the \emph{collective} experience of all the participating agents in the previous rounds is inspired by platforms such as Yelp which aggregate feedback from all users to provide a single review rating which is publicly accessible. The decision models in this paper are reminiscent of well-known models studied in the context of convergence to correlated equilibrium, e.g., \cite{Foster.Vohra:97,Fudenberg.Levine:99,Hart.Mas-Colell:00}. This is not completely surprising given that the obedience constraint has been shown to be equivalent to \emph{Bayes} correlated equilibrium~\cite{Bergemann.Morris:16}. 
Specifically, a participating agent computes regret associated with the decision in the previous round in the same spirit as \cite{Hart.Mas-Colell:00}; however, its decision in the current round depends on the cumulative regret of all the participating agents, and also, the regret determines the propensity to follow the recommendation in the current round as opposed to deviating from the decision in the previous round as in \cite{Hart.Mas-Colell:00}. The model for non-participating agents to choose routes as best response to a forecast of the flow induced by the participating agents is reminiscent of \cite{Foster.Vohra:97}. 
Our main result is that, if the signaling strategy is obedient, then the non-participating agents' forecast of the link flows induced by the participating agents converges to actual induced link flows, which in turn converges to the link flows corresponding to the Bayes correlated equilibrium associated with the signaling strategy.

The rest of the paper is organized as follows. Section~\ref{sec:model} formulates the non-atomic routing game with partial signaling, defines the notion of partial information correlated equilibrium, as well as describes the repeated game setting. Section~\ref{sec:convergence} provides convergence analysis, and Section~\ref{sec:simulation} provides illustrative simulation results. Concluding remarks are provided in Section~\ref{sec:conclusion}. The proofs of all the technical results are collected in the Appendix.

We end this section by defining key notations to be used throughout this paper. Let $\triangle(X)$ denote the set of all probability distributions on $X$. 
For a vector $x \in \RR^n$, $\diag(x)$ will denote the $n \times n$ diagonal matrix with elements of $x$ on the main diagonal. 
For an integer \( n \), we let \( [n]:= \{1, 2, \ldots, n\} \). For $\lambda \geq 0$, let $\simplex_n(\lambda):=\setdef{x \in \real_{\geq 0}^n}{\sum_{i \in [n]} x_i = \lambda}$ be the $(n-1)$-dimensional probability simplex of size $\lambda$. When $\lambda=1$, we shall simply denote the simplex as $\simplex_n$ for brevity in notation. For a real number $r \in \real$, we let $[r]^+ \:= \text{max}(r, 0)$, and $\lfloor r \rfloor$ to be the largest integer no greater than $r$.

\section{Model}
\label{sec:model}

\subsection{Non-atomic Routing Game with Partial Signaling}

Consider a network consisting of $\nlinks$ parallel links between a single source-destination pair. Without loss of generality, let the agent population generate a unit volume of traffic demand. 
The link latency functions $\congfunc_{\omega,i}(f_i)$, $i \in [\nlinks]$, give latency on link $i$ as a function of flow $f_i$ through them, conditional on the \emph{state} of the network $\state \in \allstates=\{\state_1, \ldots,\state_{\nstates}\}$. Throughout the paper, we shall make the following basic assumption on these functions.
\begin{assumption}
\label{ass:continuous}
For every $i \in [\nlinks]$, $\state \in \Omega$, $\congfunc_{\omega,i}$ is a non-negative, continuously differentiable and non-decreasing function.
\end{assumption}
At times, we shall strengthen the assumption to $\congfunc_{\omega,i}$ being strictly increasing. For Proposition~\ref{prop:homo}, Lemma~\ref{lemma:convergence}, Lemma~\ref{lemma:lipschitz} and Theorem~\ref{prop:convergence}, we restrict ourselves to general polynomial latency function in the following form:
\begin{equation}
\label{eq:polynomial-latency-function}
\congfunc_{\omega,i}(f_i) =  \sum^D_{d=0} \alpha_{d,\omega,i} \, f_i^d, \, i \in [\nlinks], \, \, \state \in \allstates
\end{equation}
with $\alpha_{0,\state,i} \geq 0$ and $\alpha_{1,\state,i} \geq 0$. We shall also let $\alpha_{d,\omega}$ refer to the $\nlinks \times 1$ column vector whose entries are $\alpha_{d,\state,i}$, and $\alpha_d$ refer to the $\nstates \times \nlinks$ matrix whose entries are $\alpha_{d,\state,i}$.

Let $\state \sim \prior \in \interior(\triangle(\allstates))$, for some prior $\prior$ which is known to all the agents. 
The agents do not have access to the realization of $\state$, but a fixed fraction $\pfrac \in [0,1]$ of the agents receives private route recommendations conditional on the realized state. These conditional recommendations are generated by a \emph{signal} $\signal=\{\signal_{\state} \in \simplex_{\nlinks}(\pfrac): \, \state \in \allstates\}$.

\begin{remark}
The signal defined above is \emph{diagonal atomic}~\cite{Zhu.Savla:TCNS20}. A generalization is $\signal=\{\signal_{\state} \in \triangle(\simplex_{\nlinks}(\pfrac)): \, \state \in \allstates\}$. We postpone consideration of such a general setting to future work. 
\end{remark}

\subsection{Repeated Game Setting}
\ksmarginphantom{what is the correct terminology: repeated game or multi-stage game?}
Consider the following repeated game setting. The strategy space of a participating agent (P-agent) is $\{\mathrm{obey}, \\ \mathrm{do \, not \, obey}\}$, whereas the strategy space of a non-participating Bayesian agent (B-agent) is $\{1, \ldots, \npaths\}$. The $\mathrm{do \, not \, obey}$ behavior of the P-agents is modeled by the row-stochastic matrix $P$ with zeros on the diagonal, where $P_{ij}$ is the fraction of $\mathrm{do \, not \, obey }$ P-agents who are recommended $i$ but choose $j$. The choice of a P-agent is driven by a notion of regret, and the decision of a B-agent is myopic best response to a forecast of P-agent decisions. The details are as follows. 
%
%
%
%

\subsubsection{Route choice model for P-agents}
\label{sec:P-dynamic}
The outline of the decision making process of P-agents is as follows. 
At the end of stage $k$, every P-agent computes difference between the payoffs associated with the recommended choice and the alternative choice(s). These payoff differences are then aggregated over the entire population of $P$-agents to give $u(k)$. The average of the initial condition $m(1)$ and $u(1), \ldots, u(k)$, denoted as $m(k+1)$, is then mapped to regret as $\theta(m(k+1)) \in [0,1]$, which equals the fraction of P-agents who do not follow recommendation in stage $k+1$. The details for each step in this process is provided next. 

We assume that upon completion of the trips, all agents ($P$- and $B$-agents) have access to traffic report from the $k$-th stage. This report consists of $\omega(k)$ and $\{\ell_{\omega(k),i}\}_{i \in [n]}$. For simplicity, first consider the two-link case with $\ell_{\omega(k),1}>\ell_{\omega(k),2}$. For a $P$-agent who was recommended route 1, irrespective of whether she obeys the recommendation or not, her recommendation is sub-optimal by $\ell_{\omega(k),1}-\ell_{\omega(k),2}$. In the general case of $n \geq 2$ links, for a $P$-agent who obeys recommendation to take route $i$, the sub-optimality of the recommendation is $\congfunc_{\omega(k),i} -\sum_{j \in [n]} P_{ij} \congfunc_{\omega(k),j}$. On the other hand, for a $P$-agent who does not obey the recommendation to take route $i$ but rather takes route $j$, the sub-optimality of her recommendation is $\congfunc_{\omega(k),i} -\congfunc_{\omega(k),j}$. Taking into account the number of $P$-agents who are recommended different routes and the fraction of them who $\mathrm{obey}$ or $\mathrm{do \, not \, obey}$ the recommendation, the aggregation of payoff difference over the entire P-agent population is given by:
\begin{align}
u(k) := & \sum_{i \in [n]} \Big(\congfunc_{\omega(k),i} -\sum_{j \in [n]} P_{ij} \congfunc_{\omega(k),j} \Big) \pi_{\omega(k),i} \left(1-\theta(m(k))\right) + \sum_{i,j \in [n]} \Big(\congfunc_{\omega(k),i} - \congfunc_{\omega(k),j} \Big) P_{ij}\pi_{\omega(k),i} \theta(m(k)) \nonumber \\
=& \sum_{i,j \in [n]} \Big(\congfunc_{\omega(k),i} - \congfunc_{\omega(k),j} \Big) P_{ij} \pi_{\omega(k),i} \nonumber \\
= & {\pi_{\omega(k)}}^T (I-P) \congfunc_{\omega(k)}
\label{eq:instant} 
\end{align}
where $\congfunc_{\omega(k)}$ is the $\nlinks \times 1$ column vectors whose entries are $\congfunc_{\omega(k),i}, \, i \in [n]$.
The average of these instantaneous payoff differences (as well as the initial condition $m(1)$) is:
\begin{equation}
\label{eq:S-agent-update}
\begin{split}
m(k+1) & = \frac{1}{k+1} \left(m(1)+u(1) + \ldots u(k) \right)
\\
& = \frac{k}{k+1} m(k) + \frac{1}{k+1} u(k)
\end{split}
\end{equation}

We adopt the following notion of regret:
\begin{equation}
\label{eq:rating-to-mistrust}
\theta(m(k))=\frac{[m(k)]^+}{m_{\max}}
\end{equation}
where $m_{\max}$ is chosen to be sufficiently large so that $\theta(m(k)) \in [0,1]$ for all possible values of $m(k)$. Since $[m(k)]^+$ can be upper bounded by $\sum_{i \in [n]} \sum_{ d=0}^D \max_{\omega \in \Omega} \alpha_{d,\omega,i}$, it can then serve as a lower bound for $m_{\max}$.

$\theta(k) = \theta(m(k))$ is interpreted as the fraction of agents who do not follow recommendation. 
Therefore, the link flows induced by $P$-agents be given by: \ksmarginphantom{need more details for \eqref{eq:p flow dynamics-m-general}}
\begin{equation*}
\label{eq:p flow dynamics-m-general}
\begin{split}
x_i(k) = x_i(m(k),\omega(k)) = \pfrac \pi_{\omega(k),i} \left(1-\theta(m(k))\right) + \pfrac \sum_{j \in [n]} P_{ji} \pi_{\omega(k),j} \theta(m(k))
\end{split}
\end{equation*}
which in matrix form becomes:
\begin{align}
x(k) = x(m(k),\omega(k)) & = \pfrac \pi_{\omega(k)} \left(1-\theta(m(k))\right) + \pfrac \sum_{j \in [n]} P_{ji} \pi_{\omega(k),j} \theta(m(k)) \nonumber \\
& = \pfrac \left(\pi^{\omega(k)} - \theta(m(k)) \pi^{\omega(k)} + \theta(m(k)) P^T \pi_{\omega(k)} \right) \nonumber \\
& = \pfrac \pi_{\omega(k)} + \pfrac \theta(m(k)) (P^T - I ) \pi_{\omega(k)}
\label{eq:p flow dynamics-m-general-matrix}
\end{align}
where $x(m(k),\omega(k))  \in \simplex_{\nlinks}(\pfrac)$.

\begin{remark}
The framework in \eqref{eq:instant}-\eqref{eq:p flow dynamics-m-general-matrix} is reminiscent of the \emph{regret matching} framework of \cite{Hart.Mas-Colell:00}. Details on comparison are provided in Section~\ref{sec:regret}. A few points are worth special emphasis:
\begin{enumerate}
\item The aggregation of payoff difference over all $P$-agents in \eqref{eq:instant} is inspired by platforms such as Yelp. 
\item In the framework of \cite{Hart.Mas-Colell:00}, $[m]^+$ is referred to as regret and $m_{\max}$ is the inertia parameter which captures the propensity of a $P$-agent to stick to the action choice in the previous round. On the other hand, in our setup, $\theta=[m]^+/m_{\max}$ is interpreted as the degree of obedience of an individual agent,  tuned by the parameter $m_{\max}$.
\end{enumerate}
\end{remark}

\subsubsection{Route choice model for B-agents}
The B-agents forecast the flow induced by P-agents for different realizations of $\omega$ 
using a model structurally similar to \eqref{eq:p flow dynamics-m-general-matrix}, as follows:
\begin{equation*}
\label{eq:estimated p flow dynamics-m-general}
\begin{aligned}
\hat{x}_i(k) = \hat{x}_i(\hat{\theta}(k),\omega(k)) = \pfrac \pi_{\omega(k),i} \left(1-\hat{\theta}(k)\right) + \pfrac \sum_{j \in [n]} P_{ji} \pi_{\omega(k),j} \hat{\theta}(k)
\end{aligned}
\end{equation*} 
or in matrix form as:
\begin{equation}
\label{eq:estimated p flow dynamics-m-general-matrix}
\hat{x}(k) = \hat{x}(\hat{\theta}(k),\omega(k)) = \pfrac \pi_{\omega(k)} + \pfrac \hat{\theta}(k) (P^T - I ) \pi_{\omega(k)}
\end{equation}
where $\hat{\theta}$ is B-agents' forecast of $\theta$, $\hat{x}(k)  \in \simplex_{\nlinks}(\pfrac)$. The \emph{simple} exponential smoothing model for this forecast can be written in the following equivalent forms:
\begin{align}
\hat{\theta}(k+1)& = \beta(k+1) \theta(k) + (1-\beta(k+1)) \hat{\theta}(k) \nonumber \\
 & = \hat{\theta}(k)+\beta(k+1) e_{\theta}(k) \label{eq:smoothing-model}
\end{align}
where $e_{\theta}(k):=\theta(k)-\hat{\theta}(k)$, $\beta(k+1) \in (\beta_{\min},\beta_{\max}) \subset (0,1)$ is the smoothing parameter, and $\hat{\theta}(1) \in [0,1]$ is the initial forecast. 


Let us denote the myopic best response to the forecast as $y(\hat{\theta}(k))$. For a given $\theta \in [0,1]$, $y(\theta)$ is the unique (assuming strictly increasing latency functions $\{\ell_{\omega,i}\}_{\omega,i}$, more details in Lemma~\ref{lemma:lipschitz}) $y \in \simplex_n(1-\pfrac)$ satisfying:
\begin{equation}
\label{eq:BWE}
\begin{split}
y_i & > 0 \implies 
\\
& E_{\omega \sim \prior}[\congfunc_{\omega,i}(\hat{x}_i(\theta,\omega)+y_i)] \leq E_{\omega \sim \prior}[\congfunc_{\omega,j}(\hat{x}_j(\theta,\omega)+ y_j)], \qquad i, j \in [n]
\end{split}
\end{equation}

\begin{remark}
\label{rem:B-forecast}
Implementation of \eqref{eq:smoothing-model} requires B-agents to have access to $\theta(k)$. As we assumed in Section~\ref{sec:P-dynamic}, $B$-agents have access to state $\omega(k)$, delays $\{\ell_{\omega(k),i}\}_{i \in [n]}$. B-agents also know the fixed signal $\pi$, as well as explicit form of latency functions. For a given $\hat{\theta}(k)$, \eqref{eq:estimated p flow dynamics-m-general-matrix} gives $\hat{x}_i(k)$. \eqref{eq:BWE} then gives $y(\hat{\theta}(k))$, i.e., the flow induced by $B$-agents is known to them. Assuming the latency functions are strictly increasing, the total link-wise flows can be inferred from other information known to $B$-agents. They can then infer the actual inflows induced by $P$-agents. Using the inverse of \eqref{eq:p flow dynamics-m-general-matrix}, $B$-agents can then get $\theta(k)$. 
\end{remark}

An equivalent variational inequality characterization of $y(\theta)$ is that it is the unique $y \in \simplex_n(1-\pfrac)$ satisfying
\begin{equation}
\label{eq:BEW-VI}
\sum_{i \in [n]} \, E_{\omega \sim \prior}[\congfunc_{\omega,i}(\hat{x}_i(\theta,\omega)+y_i)] (z_i - y_i) \geq 0 \qquad \forall \, z \in \simplex_n(1-\pfrac)
\end{equation}

In matrix form, $y(\theta)$ satisfies:
\begin{equation}
\label{eq:BEW-VI-matrix}
E_{\omega \sim \prior} \, \left[ (z-y(\theta))^T \, \congfunc_{\omega}(\hat{x}(\theta,\omega)+y)\right] \geq 0 \qquad \forall \, z \in \simplex(1-\pfrac)
\end{equation}

Note that the total link flow forecasted by B-agents is $\hat{x}(\hat{\theta}(k),\omega(k))+y(\hat{\theta}(k))$, but the actual flow is $x(m(k),\omega(k))+y(\hat{\theta}(k))$.

\begin{remark}
\label{rem:BWE-lipschitz}
\begin{enumerate}
\item One can show that the map defined in \eqref{eq:BWE} is Lipschitz continuous. Detailed proof in Lemma~\ref{lemma:lipschitz}. 
\item The forecast and best response combination of $B$ agent decisions is reminiscent of \cite{Foster.Vohra:97}, where convergence is established to correlated equilibrium set in two player games if the forecast is \emph{calibrated} to the plays of the opponent. We argue in Section~\ref{sec:forecast} that, with appropriate adaptation to the non-atomic setting of this paper, the forecast strategy of $B$-agents could also interpreted as being calibrated to $P$-agents' actions.  
\end{enumerate}
\end{remark}

\section{Convergence Analysis and Discussion}
\label{sec:convergence}
We shall establish convergence of \eqref{eq:instant}-\eqref{eq:smoothing-model} for $\signal$ which are \emph{obedient}, i.e., for which there exists $y \in \simplex_{\nlinks}(1-\pfrac)$ such that: 
\begin{subequations}
\label{eq:obedience-hetero-v1}
\begin{align}
\sum_{\state \in \Omega} \ell_{\omega,i}(\signal_{\state,i} + y_i) \, \signal_{\state,i} \, \prior(\state) & \leq \sum_{\state \in \Omega}  \ell_{\omega,j}(\signal_{\state,j} + y_j) \, \signal_{\state,i} \, \prior(\state), \quad i, j \in [\npaths] 
\label{eq:obedience-hetero-v1:obedience} \\
\sum_{\state \in \Omega}  \ell_{\omega,i}(\signal_{\state,i} + y_i) \, y_i \, \prior(\state) &
\leq \sum_{\state \in \Omega} \ell_{\omega,j}(\signal_{\state,j} + y_j) \, y_i \, \prior(\state), \quad i, \, j \in [\npaths] 
\label{eq:obedience-hetero-v1:nash}
\end{align}
\end{subequations}

Following \cite{Bergemann.Morris:16},  $(\signal,y)$ is therefore the \emph{Bayes Correlated Equilibrium} induced by $\signal$. We start the convergence analysis by considering the extreme case when there are no B-agents, i.e., when $\pfrac=1$.
\begin{proposition}
\label{prop:homo}
Consider the dynamics in \eqref{eq:instant}-\eqref{eq:p flow dynamics-m-general-matrix} with $\nu=1$, and general polynomial latency functions in \eqref{eq:polynomial-latency-function}. For any $\prb$, obedient $\pi$, and every $m(1) \in [-m_{\max},+m_{\max}]$, we have that, almost surely
$$
\lim_{k \rightarrow \infty} x_i(k) - \pi_{\omega(k),i} = 0, \qquad i \in [n]
$$
\end{proposition}

Towards consideration of the general case of heterogeneous population, we first provide an analysis of the case in the absence of the aggregator, i.e., for some fixed $m(k) \equiv m$.

\begin{proposition}
\label{prop:convergence-lambda-zero}
Consider the dynamics in \eqref{eq:instant}-\eqref{eq:smoothing-model} for some fixed $m(k) \equiv m \in [-m_{\max},+m_{\max}]$. For any $\mu_0$, obedient $\pi$, $\nu \in (0,1)$, $\hat{\theta}(1) \in [0,1]$, we have that
\begin{equation*}
\lim_{k \to \infty} \hat{\theta}(k) = \theta(m)
\end{equation*}
\end{proposition}

Next, we extend the analysis to the case when $m(k)$ evolves according to \eqref{eq:instant}-\eqref{eq:S-agent-update}.
\begin{proposition}
\label{prop:gamma-lambda-non-zero}
Consider the dynamics in \eqref{eq:instant}-\eqref{eq:smoothing-model}. For any $\mu_0$, obedient $\pi$, $\nu \in (0,1)$, $\hat{\theta}(1) \in [0,1]$, $m(1) \in [-m_{\max},+m_{\max}]$, we have that
\begin{equation*}
\lim_{k \to \infty} \hat{\theta}(k) - \theta(m(k))=0
\end{equation*}
\end{proposition}

The following results will be useful in proving convergence of link-wise flows. 

\begin{lemma}
\label{lemma:convergence}
Consider the dynamics in \eqref{eq:instant}-\eqref{eq:smoothing-model} for general polynomial latency functions in \eqref{eq:polynomial-latency-function}. For any $\mu_0$, obedient $\pi$, $\nu \in (0,1)$, and initial conditions $\hat{\theta}(1) \in [0,1]$ and $m(1) \in [-m_{\max},+m_{\max}]$, there exists a subsequence $\{m(k_s)\}_s$, whose limit is negative almost surely.
\end{lemma}

\begin{lemma}
\label{lemma:lipschitz}
The mapping $y(\theta): [0,1] \rightarrow \simplex_n(1-\pfrac)$ defined in \eqref{eq:BWE} for general polynomial latency functions in \eqref{eq:polynomial-latency-function} is Lipschitz continuous.
\end{lemma}

We are now ready to state the main result. 

\ksmarginphantom{shouldn't this theorem be about the total flow?}
\begin{theorem}
\label{prop:convergence}
Consider the dynamics in \eqref{eq:instant}-\eqref{eq:smoothing-model} for general polynomial latency functions in \eqref{eq:polynomial-latency-function}. For any $\mu_0$, obedient $\pi$, $\nu \in (0,1)$, and initial conditions $\hat{\theta}(1) \in [0,1]$ and $m(1) \in [-m_{\max},+m_{\max}]$,
we have that, almost surely
$$
\lim_{k \rightarrow \infty} x_i(k) - \nu \pi_{\omega(k),i} = 0, \qquad i \in [n]
$$
\end{theorem}

\section{Simulations}
\label{sec:simulation}
We report simulation results which suggest that the main convergence result is robust to natural variations of the dynamics in  \eqref{eq:instant}-\eqref{eq:smoothing-model}. We consider a network with two parallel links. 
Unless noted otherwise, in all the scenarios, 
%
$\mu_0(\omega_1)=0.6=1-\mu_0(\omega_2)$, $\nu=0.5$, $\theta(1)=0.5$, $\hat{\theta}(1)=0.25$, and the total demand is set to be $1$. Also, the link latency functions are affine with coefficients: 
\begin{equation*}
\alpha_0 = 
\kbordermatrix{&i=1&i=2\\
\omega_1&5&25\\
\omega_2&20&15}, \quad 
\alpha_1 = 
\kbordermatrix{&i=1&i=2\\
\omega_1&4&2\\
\omega_2&1&2}
\end{equation*} 

\begin{figure}[htb!]
\begin{center}
\begin{minipage}[c]{.45\textwidth}
\begin{center}
\includegraphics[width=1.0\textwidth]{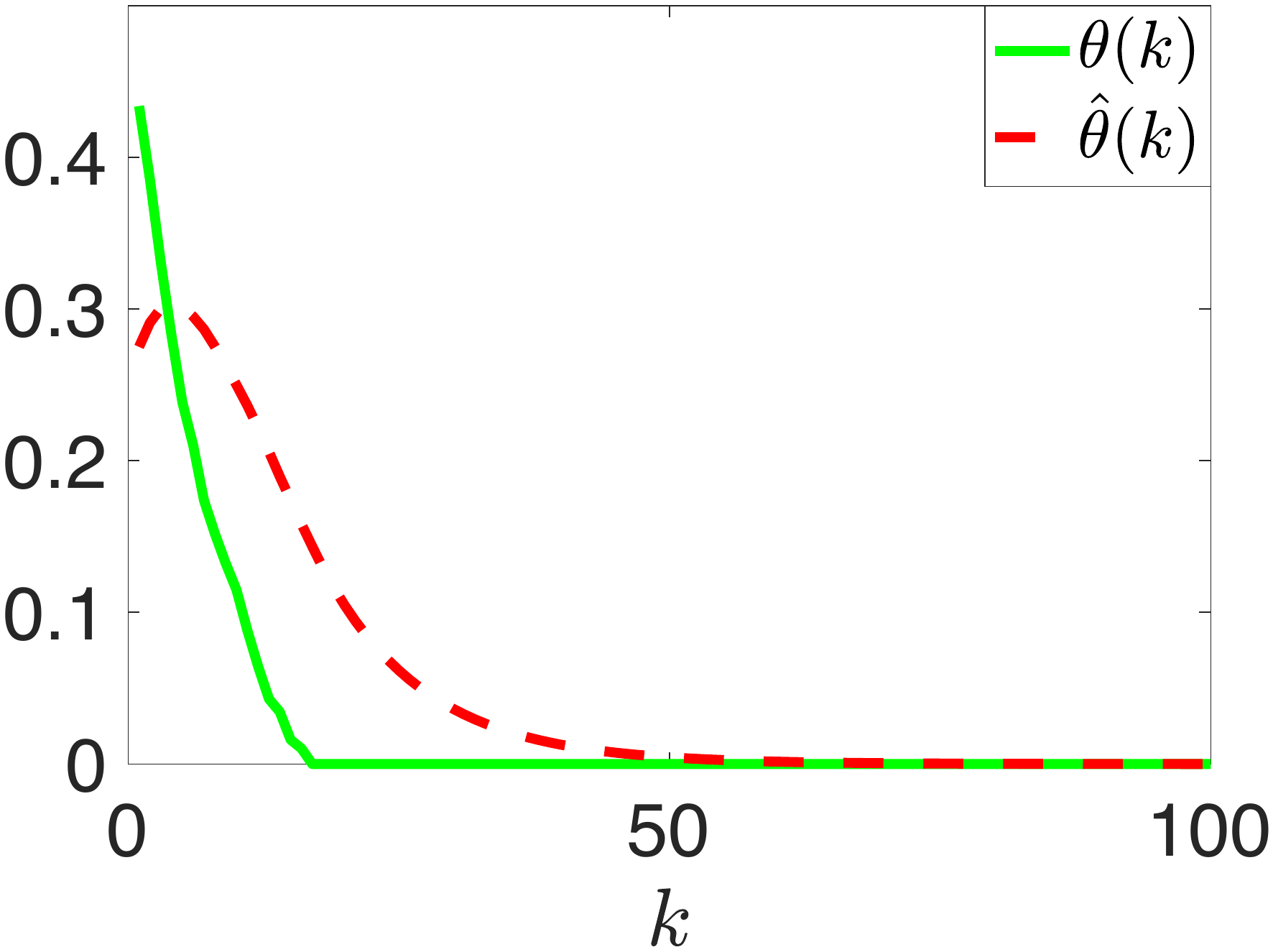}
\\
(a)
\end{center}
\end{minipage}
\begin{minipage}[c]{.45\textwidth}
\begin{center}
\includegraphics[width=1.0\textwidth]{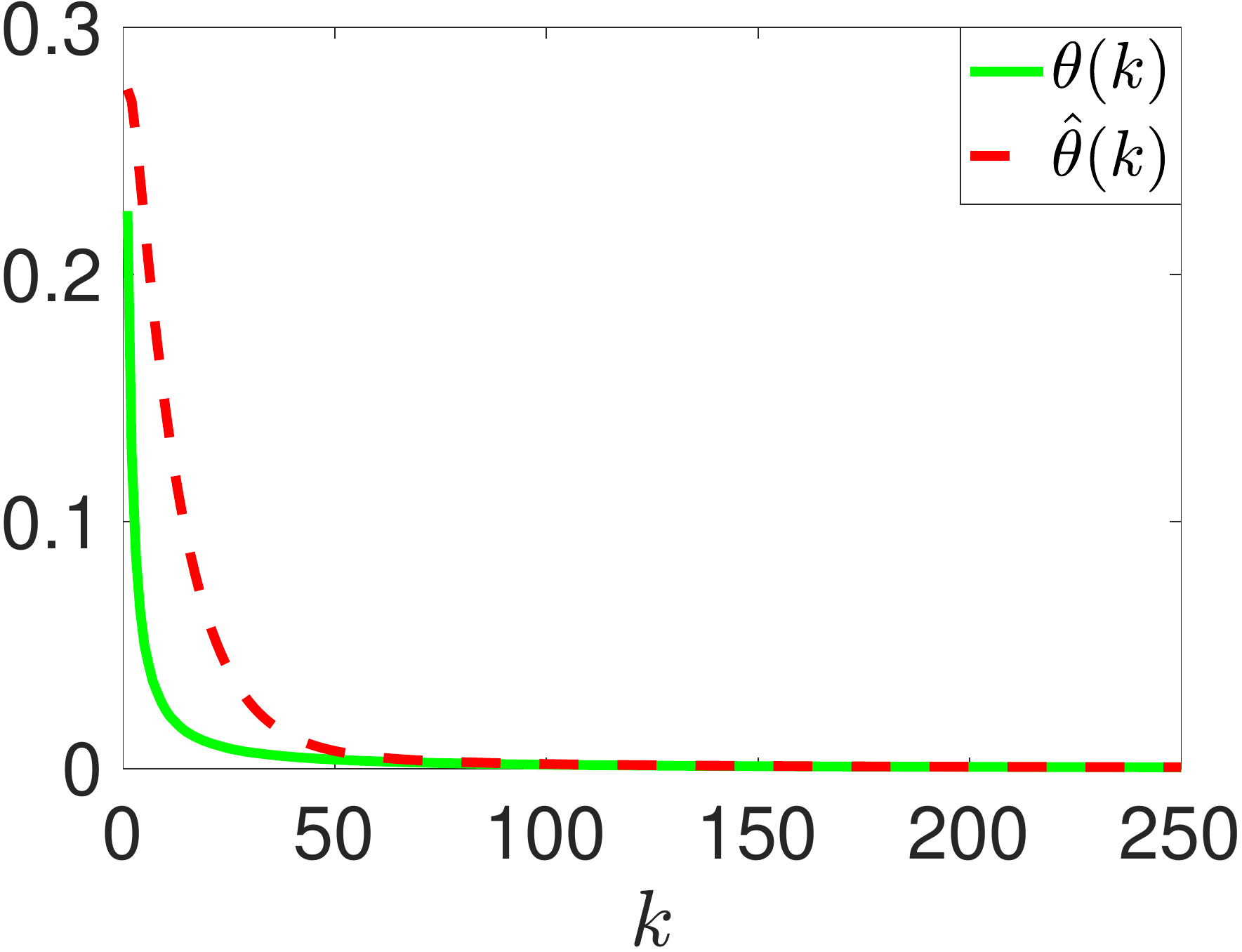}
\\
(b)
\end{center}
\end{minipage}
\end{center}
\caption{\small \sf Convergence of regret and its forecast over two parallel links for (a) review aggregation with discounting factor $\lambda=0.9$, and (b) dynamic participation rate.}
\label{fig:simulation}
\end{figure}

In Scenario (a), we replace \eqref{eq:S-agent-update} with discounted review aggregation: $m(k+1)=\lambda m(k)+(1-\lambda)u(k)$. In Scenario (b), we let the participation rate also be dynamic, and in particular, we let $\nu(k)=\theta(k)$. Figure~\ref{fig:simulation}, which shows results for these scenarios, suggests that convergence of regret forecast as well as actual regret to zero is achieved even under these variations. 
%

\section{Conclusion}
\label{sec:conclusion}

In this paper, we provided convergence analysis for a non-atomic routing game with partial signaling. The decision models are reminiscent of adaptive models studied previously in the context of dynamic procedures leading to the set of correlated equilibria. Our analysis lays the foundation to extend such analysis to repeated scenarios with explicit presence of a coordinator, albeit in the context of non-atomic routing games. 

\ksmarginphantom{need to briefly discuss the challenges}
There are several natural directions for future work. We plan to extend the analysis to general increasing latency functions, and to settings involving multiple destinations. The notion of conditional universal consistency in \cite{Fudenberg.Levine:99} potentially offers a framework to unify and generalize the specific decision models of the participating and non-participating agents considered in this paper. Finally, it would be interesting to consider a multiscale routing decision framework to enable integration of traffic flow dynamics into the decision-theoretic analysis, e.g., as in \cite{Como.Savla.ea:SICON11}.

\bibliographystyle{ieeetr}
\bibliography{ksmain,savla}

\appendix

\subsection{Connection to Calibrated Forecast}
\label{sec:forecast}
\cite{Foster.Vohra:97} studies a two player repeated game, in which, in each round, each player forecasts the probability $p=(p_1,\ldots,p_{\nlinks})$ of decisions for the other player. Convergence to the set of correlated equilibria is established if the forecasting rule is \emph{calibrated}. Let $N(p,k)$ be the number of rounds up to the $k$-th round in which the forecast is $p$, and let $\rho(p,j,k)$ be the fraction of these rounds for which the other player plays $j$. Then the forecasting rule is said to be calibrated if:
\begin{equation}
\label{eq:calibration-original}
\lim_{k \to \infty} \sum_p \Big|\rho(p,j,k)-p_j\Big| \frac{N(p,k)}{k} = 0 \qquad \forall \, j
\end{equation}

In contrast to \cite{Foster.Vohra:97}, our setting considers non-atomic agents, and the utility of every player depends also on the state $\state$, which changes from one round to the next. Non-atomicity gives $B$ agents sufficient samples to compare forecast with outcome on a per round basis. Specifically, the flow forecast $\hat{x}_j$ (after normalization by $\pfrac$) could be interpreted as the forecast of the probability with which an individual $P$-agent will pick route $j$, and similarly $x_j$ is the probability with which a $P$-agent actually picks $j$. Therefore, a reasonable adaptation of \eqref{eq:calibration-original} to our setting is that the $B$ agents' forecast is calibrated if: 
%
%
\begin{equation}
\label{eq:calibration-modified}
\lim_{k \to \infty} \frac{1}{k} \sum_{t=1}^k \Big|x_j(t) - \hat{x}_j(t) \Big| = 0 \qquad \forall \, j
\end{equation}
for almost every i.i.d. realization of $\state(t) \sim \prior$.


Proposition~\ref{prop:gamma-lambda-non-zero} shows that $\lim_{k \to \infty} \hat{\theta}(k) - \theta(m(k))=0$. Combine with \eqref{eq:p flow dynamics-m-general-matrix} and \eqref{eq:estimated p flow dynamics-m-general-matrix}, one can easily get $\lim_{k \to \infty} x_j(k) - \hat{x}_j(k)=0$ for $\forall \, j$. Thus the forecasting rule in \eqref{eq:estimated p flow dynamics-m-general-matrix}-\eqref{eq:smoothing-model} is calibrated in the sense of \eqref{eq:calibration-modified}.


\subsection{Connection to Regret-Matching}
\label{sec:regret}
\cite{Hart.Mas-Colell:00} considers the following repeated game between finite players. In round $k$, a player who chose action $i$ in round $k-1$, switches to $j \neq i$ with probability $\frac{[m_{i \to j}(k)]^+}{m_{\max}}$, and sticks to $i$ with probability $1-\sum_{j\neq i}\frac{[m_{i \to j}(k)]^+}{m_{\max}}$, where $m_{i \to j}(k)$ is a measure of \emph{regret} for not having played $j$ every time that $i$ was played in the past, and the \emph{inertia} parameter $m_{\max}$ is sufficiently large to ensure that the probabilities are well defined. Formally, the regret is computed as: $m_{i \to j}(k):=\frac{1}{k-1}\sum_{t=1: \, \text{player chooses }i}^{k-1} (U(j, s^{-}(t))-U(s(t)))$, where $U$ is the utility function of the player under consideration, $s(t)$ is the set of actions chosen by all the players in round $t$, and $s^{-}(t)$ is the set of actions of all the players except the one under consideration. 


\ksmarginphantom{does swapping the time average and population average sequence lead to the same model?}
The similarities and differences with the decision model of the $P$ agents in our setup is now apparent. An individual $P$ agent computes regret for its action to follow or to not to follow the recommendation in the previous round in the same spirit as \cite{Hart.Mas-Colell:00}. However, this regret is not categorized in terms of the specific action taken in the previous round. Such an absence of conditioning facilitates efficient aggregation of regrets of all the $P$ agents as a simple summation. The time-averaging of regrets is achieved through \eqref{eq:S-agent-update}, where $m_{\max}$ plays the role as the inertia parameter similar to \cite{Hart.Mas-Colell:00}.

\subsection{Proof of Proposition~\ref{prop:homo}}
Plugging \eqref{eq:instant} into \eqref{eq:S-agent-update} and substituting \eqref{eq:p flow dynamics-m-general-matrix} gives
\begin{align}
m(k) = & \frac{m(1)}{k}+\frac{1}{k} \sum_{t=1}^{k-1} u(t) \nonumber \\
=&\frac{m(1)}{k}+\frac{1}{k} \sum_{t=1}^{k-1} {\pi_{\omega(t)}}^T (I-P) \congfunc_{\omega(t)}(x(t)) \nonumber \\
=&\frac{m(1)}{k}+\frac{1}{k} \sum_{t=1}^{k-1} {\pi_{\omega(t)}}^T (I-P) \left(\alpha_{0,\omega(t)} + \sum_{d=1}^D \diag(\alpha_{d,\omega(t)}) \diag(x(t))^{d-1} x(t)\right) \nonumber \\
\label{eq:mistrust-m1-general}
= & \frac{m(1)}{k}+\frac{1}{k} \sum_{t=1}^{k-1} {\pi_{\omega(t)}}^T (I-P) \left(\ell_{\omega(t)}(\pi_{\omega(t)}) + \sum_{d=1}^D \diag(\alpha_{d,\omega(t)}) \left(\diag(x(t))^{d-1} x(t) - \diag(\pi_{\omega(t)})^{d-1} \pi_{\omega(t)}\right) \right) \nonumber \\
= & \frac{m(1)}{k}+\frac{1}{k} \sum_{t=1}^{k-1} {\pi_{\omega(t)}}^T (I-P) \ell_{\omega(t)}(\pi_{\omega(t)}) \nonumber \\
& \qquad \qquad \quad - \frac{1}{k} \sum_{t=1}^{k-1} \theta(t) \underbrace{{\pi_{\omega(t)}}^T (I-P) \sum_{d=1}^D \diag(\alpha_{d,\omega(t)}) \sum_{i=0}^{d-1} \diag(x(t))^{d-1-i} \diag(\pi_{\omega(t)})^{i}  (I-P^T) \pi_{\omega(t)}}_{\geq 0} \nonumber \\
\leq &  \frac{m(1)}{k}+\frac{1}{k} \sum_{t=1}^{k-1} {\pi_{\omega(t)}}^T (I-P) \congfunc_{\omega(t)}(\pi_{\omega(t)}) \nonumber \\
= &  \frac{m(1)}{k}+\sum_{i, j \in [n]} P_{ij} \frac{1}{k} \sum_{t=1}^{k-1} \pi_{\omega(t),i} \left(\congfunc_{\omega(t),i}(\pi_{\omega(t)})-\congfunc_{\omega(t),j}(\pi_{\omega(t)})\right)
\end{align}
\eqref{eq:obedience-hetero-v1} and the strong law of large numbers implies that, almost surely, 
$\lim_{k \to \infty}\frac{1}{k} \sum_{t=1}^{k-1} \pi_{\omega(t),i} \\ \left(\congfunc_{\omega(t),i}(\pi_{\omega(t)}) -\congfunc_{\omega(t),j}(\pi_{\omega(t)})\right) \leq 0$ for all $i, j \in [n]$, and therefore, almost surely, $\limsup_{k \to \infty} m(k) \leq 0$, and hence $\lim_{k \to \infty} \theta(k)=0$ from \eqref{eq:rating-to-mistrust}. The proposition then follows from \eqref{eq:p flow dynamics-m-general-matrix}. 

\subsection{Proof of Proposition~\ref{prop:convergence-lambda-zero}}
\eqref{eq:smoothing-model} implies $e_{\theta}(k+1) = (1-\beta(k+1)) e_{\theta}(k)$, and hence $e_{\theta}(k+1)=\prod_{t=2}^{k+1} (1-\beta(t)) e_{\theta}(1)$. The proposition then follows from the fact that $(1-\beta_{\max})^k e_{\theta}(1) \leq e_{\theta}(k+1) \leq (1-\beta_{\min})^k e_{\theta}(1)$.

\subsection{Proof of Proposition~\ref{prop:gamma-lambda-non-zero}}
\eqref{eq:smoothing-model} implies
\begin{equation}
\label{eq:P-dynamic}
\begin{aligned}
e_{\theta}(k+1) = (1- \beta(k+1)) e_{\theta}(k)+\delta(k)
\end{aligned}
\end{equation}
where 
$\delta(k)=\theta(k+1)-\theta(k)$. Using the expression of $\theta$ from \eqref{eq:rating-to-mistrust} and noting that $\left|[a]^+-[b]^+\right|\leq[a-b]^+$ for all $a, b \in \mathbb{R}$, we get
\begin{align*}
\frac{-[u(k)-m(k)]^+}{(k+1) m_{\max}} &\leq \delta(k) \leq \frac{[u(k)-m(k)]^+}{(k+1) m_{\max}}
\end{align*}
Since $[a-b]^+\leq|a|+|b|$ for all $a, b \in \mathbb{R}$, and both $m(k)$ and $u(k)$ are absolutely upper bounded by $m_{\max}$,  
\begin{equation}
\label{eq:harmonic-bound}
\frac{-2}{k+1} \leq \delta(k) \leq \frac{2}{k+1}
\end{equation}
Since $\delta(k)$ can only be bounded by a harmonic sequence whose partial sum sequence diverges, it is not easy to prove that $\lim_{k \to \infty} e_{\theta}(k)=0$ directly using \eqref{eq:P-dynamic}. Instead, we lower and upper bound $e(k)$ by $\underline{e}_{\theta}(k)$ and $\overline{e}_{\theta}(k)$, respectively, and show that these converge to zero. \eqref{eq:P-dynamic} can be rewritten as
\begin{equation}
\label{eq:e-explicit}
e_{\theta}(k)=\left(\prod_{t=2}^{k} (1- \beta(t))\right) e_{\theta}(1)+ \sum_{t=1}^{k-2} \left(\prod_{\tau=t+2}^{k} (1- \beta(\tau))\right) \delta(t) + \delta(k-1)
\end{equation}
Substituting \eqref{eq:harmonic-bound} into \eqref{eq:e-explicit},
\begin{equation}
\label{eq:e-bound}
\underline{e}_{\theta}(k) := \left(\prod_{t=2}^{k} (1- \beta(t))\right) e_{\theta}(1)-2 \tilde{\delta}(k) \leq e_{\theta}(k) \leq \left(\prod_{t=2}^{k} (1- \beta(t))\right) e_{\theta}(1)+2 \tilde{\delta}(k)=:\overline{e}_{\theta}(k)
\end{equation}
where $\tilde{\delta}(k)=\sum_{t=2}^{k} (1- \beta_{\min})^{k-t} \frac{1}{t}= \gamma^{k}\sum_{t=2}^{k} \frac{\gamma^{-t}}{t}$, with $\gamma:=1-\beta_{\min} \in (0,1)$. $\frac{\gamma^{-t}}{t}$ is decreasing over $(0,-\frac{1}{\ln \gamma}]$, and increasing over $[-\frac{1}{\ln \gamma},+\infty)$. Therefore, with $k^*:=\lfloor -\frac{1}{\ln \gamma}\rfloor$,
\begin{equation*}
\begin{split}
\tilde{\delta}(k) & = \gamma^k \left(\sum_{t=2}^{k^*} \frac{\gamma^{-t}}{t} + \sum_{t=k^*+1}^k \frac{\gamma^{-t}}{t}\right) \leq \gamma^k \left( \int_1^{k^*} \frac{\gamma^{-t}}{t}  \de t + \int_{k^*+1}^{k+1} \frac{\gamma^{-t}}{t}  \de t\right) \leq \gamma^k \underbrace{\int_1^{k+1} \frac{\gamma^{-t}}{t}  \de t}_{=:I_{k+1}}
\end{split}
\end{equation*}
\ksmarginphantom{need a reference}
Change of variable $z=\gamma^{-t}$ gives
$I_{k+1} = \int_{\gamma^{-1}}^{\gamma^{-(k+1)}} \frac{\de z}{\ln z}=\mathrm{Li}(\gamma^{-(k+1)})-\mathrm{Li}(\gamma^{-1})$, where $\mathrm{Li}(\cdot)$ is the logarithmic integral function \cite{abramowitz1964} whose asymptotic behavior is $\mathrm{Li}(r)=O(\frac{r}{\ln r})$. 
%
%
Since $\gamma \in (0,1)$, $\gamma^{-(k+1)} \to \infty$ as $k \to \infty$. Therefore, as $k \to \infty$,
\begin{equation*}
\begin{aligned}
0<\tilde{\delta}(k)\leq \gamma^k \left(\mathrm{Li}(\gamma^{-(k+1)})-\mathrm{Li}(\gamma^{-1})\right)&=\gamma^k \left(O\left(\frac{\gamma^{-(k+1)}}{-(k+1) \ln \gamma}\right)-\mathrm{Li}(\gamma^{-1})\right) \\
&=O\left(\frac{\gamma^{-1}}{-(k+1) \ln \gamma}\right)-\gamma^k \mathrm{Li}(\gamma^{-1}) \to 0
\end{aligned}
\end{equation*}
which leads to $\lim_{k \to \infty} \tilde{\delta}(k) =0$. Using it in \eqref{eq:e-bound} therefore gives $\lim_{k \to \infty}\underline{e}_{\theta}(k)=0$ and $\lim_{k \to \infty}\overline{e}_{\theta}(k)=0$, and hence $\lim_{k \to \infty} e_{\theta}(k)=0$.

\subsection{Proof of Lemma~\ref{lemma:convergence}}
Plugging \eqref{eq:instant} into \eqref{eq:S-agent-update} and substituting \eqref{eq:p flow dynamics-m-general-matrix} gives the following:
\begin{align}
m(k) = & \frac{m(1)}{k}+\frac{1}{k} \sum_{t=1}^{k-1} {\pi(t)}^T (I-P) \left(\alpha_{0,\omega(t)} + \sum_{d=1}^D \diag(\alpha_{d,\omega(t)}) \diag(x(t)+y(\hat{\theta}(t)))^{d-1} \left(x(t)+y(\hat{\theta}(t))\right)\right) \nonumber \\
\label{eq:m-split-general}
= & \frac{m(1)}{k}+\frac{1}{k} \sum_{t=1}^{k-1} {\pi(t)}^T (I-P) \left(\alpha_{0,\omega(t)} + \sum_{d=1}^D \diag(\alpha_{d,\omega(t)}) \diag(\nu \pi_{\omega(t)}+y(0))^{d-1} \left(\nu \pi_{\omega(t)}+y(0)\right)\right) \nonumber \\
& + \frac{1}{k} \sum_{t=1}^{k-1} {\pi(t)}^T (I-P) \sum_{d=1}^D \diag(\alpha_{d,\omega(t)}) \Bigg(\diag(x(t)+y(\hat{\theta}(t)))^{d-1} \left(x(t)+y(\hat{\theta}(t))\right) \nonumber \\
& \qquad \qquad \qquad \qquad \qquad \qquad \qquad \qquad \qquad \qquad \qquad - \diag(\nu \pi_{\omega(t)}+y(0))^{d-1} \left(\nu \pi_{\omega(t)}+y(0)\right)\Bigg) \nonumber \\
= & \frac{m(1)}{k}+\frac{1}{k} \sum_{t=1}^{k-1} {\pi(t)}^T (I-P) \left(\alpha_{0,\omega(t)} + \sum_{d=1}^D \diag(\alpha_{d,\omega(t)}) \diag(\nu \pi_{\omega(t)}+y(0))^{d-1} \left(\nu \pi_{\omega(t)}+y(0)\right)\right) \nonumber \\
& + \frac{1}{k} \sum_{t=1}^{k-1} {\pi(t)}^T (I-P) \sum_{d=1}^D \diag(\alpha_{d,\omega(t)}) \sum_{i=0}^{d-1} \diag(x(t)+y(\hat{\theta}(t)))^{d-1-i} \diag(\nu \pi_{\omega(t)}+y(0))^{i} \cdot \nonumber \\
& \qquad \qquad \qquad \qquad \qquad \qquad \qquad \qquad \qquad \qquad \qquad \qquad \left(\nu \theta(t)(P^T-I) \pi_{\omega(t)}+y(\hat{\theta}(t))-y(0)\right)
\end{align}

Since $m(k)$ is a bounded sequence, it contains a convergent subsequence $\{m((k_s)\}_s$ with limit, say, $m$. Considering \eqref{eq:m-split-general} for this subsequence, using Proposition~\ref{prop:gamma-lambda-non-zero} we have, almost surely, 
\ksmarginphantom{need to check this step carefully}
\begin{align}
\label{eq:m-split-general-limit}
m = & \underbrace{\sum_{\omega \in \Omega} \prior(\omega) {\pi_{\omega}}^T (I-P) \left(\alpha_{0,\omega} + \sum_{d=1}^D \diag(\alpha_{d,\omega}) \diag(\nu \pi_{\omega}+y(0))^{d-1} \left(\nu \pi_{\omega}+y(0)\right)\right)}_{=:m_1} \nonumber \\
& + \underbrace{\sum_{\omega \in \Omega} \prior(\omega)  {\pi_{\omega}}^T (I-P) \sum_{d=1}^D \diag(\alpha_{d,\omega}) \sum_{i=0}^{d-1} \diag(x(m, \pi_{\omega})+y(\theta(m)))^{d-1-i} \diag(\nu \pi_{\omega}+y(0))^{i} \cdot}_{=:m_2} \nonumber \\
& \quad \qquad \qquad \qquad \qquad \qquad \qquad \qquad \qquad \qquad \qquad \qquad \qquad \underbrace{\left(\nu \theta(m)(P^T-I) \pi_{\omega}+y(\theta(m))-y(0)\right)}_{=:m_2}
\end{align}
i.e., the limit $m$ of every convergent subsequence of $\{m(k)\}_k$ has to satisfy \eqref{eq:m-split-general-limit}. We now show that \eqref{eq:m-split-general-limit} admits a unique solution in $m$, and that this solution is non-positive. Towards that purpose, the following properties of $m_1$ and $m_2$ defined in \eqref{eq:m-split-general-limit} will be useful:
\begin{enumerate}
\item[(a)] $m_1 \leq 0$ for all $m$. This is due to the obedience condition;
\item[(b)] $m_2=0$ for all $m \leq 0$. This follows from the definition of $m_2$, where $\theta(m)=0$ for $m \leq 0$ from \eqref{eq:rating-to-mistrust};
\item[(c)] $m_2 \leq 0$ for all $m>0$, as follows. \eqref{eq:BEW-VI-matrix} for $y(\theta(m))$ and $y(0)$, respectively, give:
\end{enumerate} 
\begin{align}
&\sum_{\omega \in \Omega} \prior(\omega) (y(0)-y(\theta(m)))^T \Bigg(\alpha_{0,\omega} + \sum_{d=1}^D \diag(\alpha_{d,\omega}) \diag(x(m, \pi_{\omega})+y(\theta(m)))^{d-1} \left(x(m, \pi_{\omega})+y(\theta(m))\right)\Bigg) \geq 0 \nonumber \\
&\sum_{\omega \in \Omega} \prior(\omega) (y(\theta(m))-y(0))^T \Bigg(\alpha_{0,\omega} + \sum_{d=1}^D \diag(\alpha_{d,\omega}) \diag(\pfrac \signal_{\omega} + y(0))^{d-1} \left(\pfrac \signal_{\omega} + y(0)\right)\Bigg) \geq 0 \nonumber
\end{align}

Adding the two expressions gives
\begin{align}
\label{eq:expr-add}
&\sum_{\omega \in \Omega} \prior(\omega)  (y(0)-y(\theta(m)))^T \sum_{d=1}^D \diag(\alpha_{d,\omega}) \Bigg(\diag(x(m, \pi_{\omega})+y(\theta(m)))^{d-1} \left(x(m, \pi_{\omega})+y(\theta(m))\right) \nonumber \\
& - \diag(\pfrac \signal_{\omega} + y(0))^{d-1} \left(\pfrac \signal_{\omega} + y(0)\right)\Bigg) \nonumber \\
= & \sum_{\omega \in \Omega} \prior(\omega)  (y(0)-y(\theta(m)))^T \sum_{d=1}^D \diag(\alpha_{d,\omega}) \sum_{i=0}^{d-1} \diag(x(m, \pi_{\omega})+y(\theta(m)))^{d-1-i} \diag(\nu \pi_{\omega}+y(0))^{i} \cdot \nonumber \\
& \quad \qquad \qquad \qquad \qquad \qquad \qquad \qquad \qquad \qquad \qquad \qquad \qquad \left(\nu \theta(m)(P^T-I) \pi_{\omega}+y(\theta(m))-y(0)\right) \geq 0
\end{align}

Finally, 
\begin{align}
m_2 =&  -\frac{1}{\pfrac \theta(m)} \underbrace{\sum_{\omega \in \Omega} \prior(\omega) \left(\pfrac \theta(m) {\pi_{\omega}}^T (P -I)  + y^T(\theta(m)) - y^T(0)\right)  \sum_{d=1}^D \diag(\alpha_{d,\omega}) \cdot}_{=:m_3 \geq 0} \nonumber \\
& \underbrace{\sum_{i=0}^{d-1} \diag(x(m, \pi_{\omega})+y(\theta(m)))^{d-1-i} \diag(\nu \pi_{\omega}+y(0))^{i} \left(\nu \theta(m)(P^T-I) \pi_{\omega}+y(\theta(m))-y(0)\right)}_{=:m_3 \geq 0} \nonumber\\
&-\frac{1}{\pfrac \theta(m)} \underbrace{\sum_{\omega \in \Omega} \prior(\omega)  (y(0)-y(\theta(m)))^T \sum_{d=1}^D \diag(\alpha_{d,\omega})  \cdot}_{=:m_4 \geq 0 \text{ from \eqref{eq:expr-add}}} \nonumber \\
& \underbrace{\sum_{i=0}^{d-1} \diag(x(m, \pi_{\omega})+y(\theta(m)))^{d-1-i} \diag(\nu \pi_{\omega}+y(0))^{i} \left(\pfrac \theta(m)  (P^T -I) \pi^{\omega} + y(\theta(m)) - y(0)\right)}_{=:m_4 \geq 0 \text{ from \eqref{eq:expr-add}}} \nonumber
\end{align}
where we note that $\theta(m)>0$ for $m>0$.

(a), (b) and (c) imply that the right hand side of \eqref{eq:m-split-general-limit} is always non-positive. (a) and (b) also imply that the only non-positive solution to \eqref{eq:m-split-general-limit} is $m = m_1$.

\subsection{Proof of Lemma~\ref{lemma:lipschitz}}
Along the lines of Theorem 1 in \cite{Dempe1997}, we just need to verify the following three conditions: Mangasarian-Fromowitz constraint qualification (MF), strong sufficient optimality condition of second order (SOC), and constant rank constraint qualification (CR). Plugging in \eqref{eq:polynomial-latency-function} and rewriting the latency function in \eqref{eq:BWE} gives:
\begin{align}
\label{eq:BWE-latency}
\ell_{\omega,i}(\hat{x}_i(\theta,\omega)+y_i) = \sum_{d=0}^D \alpha_{d,\omega,i} (\nu \pi_{\omega.i} + \nu \theta P_i^T \pi_{\omega} - \nu \theta \pi_{\omega,i} + y_i) = \alpha_{D,\omega,i} y_i^D + \sum_{d=0}^{D-1} \beta_{d,\omega,i} y_i^d
\end{align}
where $P_i \in \real^n$ is the $i$-th column of the row-stochastic matrix $P$, and $\beta_{d,\omega,i}(\{\alpha_{k,\omega,i}\}_{k \in \{0,1,\cdots,D\}}, \pi_{\omega},\theta,P_i)$ is the coefficient of monomial of degree $d$. Taking expectation of \eqref{eq:BWE-latency} gives:
 \begin{align}
E_{\omega \sim \prior} [\ell_{\omega,i}(\hat{x}_i(\theta,\omega)+y_i)] = E_{\omega \sim \prior} [\alpha_{D,\omega,i} y_i^D + \sum_{d=0}^{D-1} \beta_{d,\omega,i} y_i^d]=\bar{\alpha}_{D,i}y_i^D + \sum_{d=0}^{D-1} \bar{\beta}_{d,i} y_i^d \nonumber
\end{align}
where $\bar{\alpha}_{D,i}:=E_{\omega \sim \prior} [\alpha_{D,\omega,i}], \bar{\beta}_{d,i}:=E_{\omega \sim \prior} [\beta_{d,\omega,i}]$. $y(\theta)$ satisfies \eqref{eq:BWE} if and only if it solves the following convex problem:
\begin{equation}
\label{prob:BWE}
\begin{aligned}
\min_{y \in \simplex_n(1-\nu)} &f(y,\theta)=\sum_{i \in [n]} \int_0^{y_i} E_{\omega \sim \prior} [\ell_{\omega,i}(\hat{x}_i(\theta,\omega)+s)] ds = \sum_{i \in [n]} \left(\frac{\bar{\alpha}_{D,i}}{D+1}y_i^{D+1} + \sum_{d=0}^{D-1} \frac{\bar{\beta}_{d,i}}{d+1} y_i^{d+1}\right) \\
\text{s.t.} \quad &g_i(y,\theta) = -y_i \leq 0, \, i \in [n] \\
&h(y,\theta)=\sum_{i=1}^n y_i -1+\nu
\end{aligned}
\end{equation}
Moreover, such $y$ is unique if $\{\ell_{\omega,i}\}_{\omega,i}$ are strictly increasing over $[0,1]$. Let the Lagrange function be $L(y,\theta,\lambda, \mu)=f(y,\theta)-\sum_{i \in [n]} \lambda_i y_i + \mu(\sum_{i=1}^n y_i -1+\nu)$. For a given $\theta^0$, $y=y^0$ is a locally optimal solution for \eqref{prob:BWE} at $\theta=\theta^0$. Now, we want to show that $y(\theta)$ is Lipschitz continuous by verifying aforementioned three conditions. 

(MF) From \eqref{prob:BWE} we have:
\begin{align}
\nabla_y g_i(y^0, \theta^0) = \begin{bmatrix}0\\ \vdots \\ -1\\ \vdots \\0\end{bmatrix}_{n \times 1}\leftarrow \text{i-th entry} \qquad \nabla_y h(y^0, \theta^0) = \begin{bmatrix}1\\1\\ \vdots \\ 1\end{bmatrix}_{n \times 1} \nonumber
\end{align}
Obviously, $\nabla_y h(y^0, \theta^0)$ is linearly independent. Let $v \in \real^n$ be such that:
\begin{equation*}
v_i=
\begin{cases}
1&\text{if} \, \, g_i(y^0, \theta^0)=0 \\
\frac{\sum_{i \in [n]}\mathbbm{1}(g_i(y^0, \theta^0)=0)}{\sum_{i \in [n]}\mathbbm{1}(g_i(y^0, \theta^0)=0)-n}&\text{if} \, \, g_i(y^0, \theta^0)\neq 0
\end{cases}
\end{equation*}
$v$ satisfies that:
\begin{align}
&\nabla_y g_i(y^0, \theta^0)^T v < 0, \quad \text{for} \, \, g_i(y^0, \theta^0)=0 \nonumber \\
&\nabla_y h(y^0, \theta^0)^T v = 0 \nonumber
\end{align}

(SOC) Let $\{H_{ij}\}_{i,j \in [n]}:=\nabla^2_{yy} L(y^0,\theta^0,\lambda, \mu)$, from \eqref{prob:BWE} we have:
\begin{equation*}
H_{ij}=
\begin{cases}
0&\text{if} \, \, i \neq j \\
D \bar{\alpha}_{D,i} y_i^{D-1}+\sum_{d=1}^{D-1} d \bar{\beta}_{d,i} y_i^{d-1}&\text{if} \, \, i=j
\end{cases}
\end{equation*}
which implies that for $\forall v \in \real^n, \, v \neq0$, we have $v^T \nabla^2_{yy} L(y^0,\theta^0,\lambda, \mu) v > 0$.

(CR) Let $I_0:=\{i | g_i(y^0, \theta^0) = 0\}$, $I$ is arbitrary but fixed subset $I \subseteq I_0$. Note that, $\sum_{i \in [n]} \mathbbm{1}(g_i(y^0, \theta^0)=0) \leq n-1$, therefore the set of gradients $\{\nabla_y g_i(y^0, \theta^0), \, i \in I\} \cup \{\nabla_y h(y^0, \theta^0)\}$ has constant rank $|I|+1$, for all $(y,\theta)$ in open neighborhood of $(y^0,\theta^0)$.

\subsection{Proof of Theorem~\ref{prop:convergence}}
Following \eqref{eq:p flow dynamics-m-general-matrix}, it suffices to show that $\lim_{k \to \infty} \theta(m(k))=0$, almost surely.

Lemma~\ref{lemma:convergence} shows that there exists a convergent subsequence $\{m(k_s)\}_s$ with limit $m \leq 0$, almost surely. That is, there exists $\tilde{s}_1>0$ such that 
\begin{equation}
\label{eq:eps1}
|m(k_s)-m|\leq -m/3 \quad \forall  s>\tilde{s}_1 \quad \text{a.s.}
\end{equation}

We now show that all the terms of the parent sequence $m(k)$ between two consecutive subsequence terms $m(k_s)$ and $m(k_{s+1})$, for $s>\tilde{s}_1$, are negative.
\eqref{eq:S-agent-update} implies, for all $s > \tilde{s}_1$,  
\begin{equation*}
m(k_s+1) = \frac{k_s}{k_s+1} m(k_s) + \frac{1}{k_s+1} u_0(k_s) + \frac{1}{k_s+1} \triangle u(k_s) 
\end{equation*}
i.e., 
\begin{equation}
\label{eq:trust-update-subsequence-delta}
m(k_s+1) - m = \frac{k_s}{k_s+1} \left(m(k_s) -m \right)+ \frac{1}{k_s+1} \left(u_0(k_s) - m \right)+ \frac{1}{k_s+1} \triangle u(k_s) 
\end{equation}
where 
\begin{equation*}
u_0(k_s):={\pi_{\omega(k_s)}}^T (I-P) \left(\alpha_{0,\omega(k_s)} + \sum_{d=1}^D \diag(\alpha_{d,\omega(k_s)}) \diag(\pfrac \signal_{\omega(k_s)} + y(0))^{d-1} \left(\pfrac \signal_{\omega(k_s)} + y(0)\right)\right)
\end{equation*}
be the instantaneous mistrust assuming B-agents estimate that P-agents are obedient, i.e., assuming $\hat{\theta}(k_s)=0$, and 
\begin{align}
& \triangle u(k_s) :=  u(k_s) - u_0(k_s) \nonumber \\
 = & {\pi_{\omega(k_s)}}^T (I-P) \sum_{d=1}^D \diag(\alpha_{d,\omega(k_s)}) \Bigg(\diag(x(k_s)+y(\hat{\theta}(k_s)))^{d-1} \left(x(k_s)+y(\hat{\theta}(k_s))\right) \nonumber \\
 & - \diag(\pfrac \signal_{\omega(k_s)} + y(0))^{d-1} \left(\pfrac \signal_{\omega(k_s)} + y(0)\right)\Bigg) \nonumber \\
 = & {\pi_{\omega(k_s)}}^T (I-P) \sum_{d=1}^D \diag(\alpha_{d,\omega(k_s)}) \sum_{i=0}^{d-1} \diag(x(k_s)+y(\hat{\theta}(k_s)))^{d-1-i} \diag(\nu \pi_{\omega(k_s)}+y(0))^{i} \left(y(\hat{\theta}(k_s))-y(0)\right)
\label{eq:delta-u}
\end{align}
In the above equations, we utilize \eqref{eq:eps1} to get $m(k_s)\leq 0$, and hence $\theta(m(k_s))=0$, for $s > \tilde{s}_1$. Using the Lipschitz property of $y(.)$ (cf. Remark~\ref{rem:BWE-lipschitz}) with \eqref{eq:delta-u} gives
\begin{equation}
\label{eq:delta-u-lipschitz}
|\triangle u(k_s)| \leq L |\hat{\theta}(k_s)|, \qquad s > \tilde{s}_1
\end{equation}
for some $L>0$. Proposition~\ref{prop:gamma-lambda-non-zero} implies that there exists $\tilde{k}_1>0$ such that $|\theta(m(k))-\hat{\theta}(k)| \leq - \frac{m}{3L}$ for all $k > \tilde{k}_1$. Let $\tilde{s}_2$ be the smallest integer greater than $\tilde{s}_1$ such that $k_{\tilde{s}_2}>\tilde{k}_1$. Therefore, $|\theta(m(k_s))-\hat{\theta}(k_s)|=|\hat{\theta}(k_s)| \leq - \frac{m}{3L}$ for all $s > \tilde{s}_2$. Combining with \eqref{eq:delta-u-lipschitz} gives
\begin{equation}
\label{eq:eps2}
|\triangle u(k_s)| \leq - \frac{m}{3}, \qquad s > \tilde{s}_2
\end{equation}

Strong law of large numbers gives $\lim_{\tau \to \infty}\left(\frac{\sum_{t=1}^{\tau}u_0(t)}{\tau}\right)=m$ almost surely, i.e., there exists $\tilde{k}_2$ such that $\left|\frac{\sum_{t=1}^{\tau} u_0(t)}{\tau} - m\right| \leq -m/6$ for all $\tau > \tilde{k}_2$. 
Therefore, for all $\tau > \tilde{k}_2$ and a non-negative integer $r$,
\begin{equation*}
\begin{split}
\tau \frac{7m}{6} \leq  \sum_{t=1}^{\tau} u_0(t) \leq \tau \frac{5m}{6}, \qquad 
(\tau+1+r) \frac{7m}{6} \leq & \sum_{t=1}^{\tau+1+r} u_0(t) \leq (\tau+1+r) \frac{5m}{6}
\end{split}
\end{equation*}
Subtraction gives
\begin{equation}
\label{eq:eps3}
|\sum_{t=\tau+1}^{\tau+1+r}(u_0(t)-m)| \leq -(2\tau+r+1) \frac{m}{6}, \qquad  \tau > \tilde{k}_2,  r \geq 0
\end{equation} 

Let $\tilde{s}$ be the smallest integer greater than $\tilde{s}_2$ such that $k_{\tilde{s}}>\tilde{k}_2$. Using \eqref{eq:eps1}, \eqref{eq:eps2} and \eqref{eq:eps3} ($\tau=k_s-1$, $r=0$) in \eqref{eq:trust-update-subsequence-delta} gives, for all $s > \tilde{s}$,
\begin{equation*}
|m(k_s+1)-m| \leq - \frac{k_s}{k_s+1} \frac{m}{3} - \frac{(2 k_s-1)}{k_s+1} \frac{m}{6}- \frac{1}{k_s+1}\frac{m}{3} < - \frac{2m}{3}
\end{equation*}
which implies that $m(k_s+1) < \frac{m}{3} \leq 0$. 

Hereafter, we show that $m(k_s+2), m(k_s+3), \ldots m(k_{s+1}-1)$ are all negative, for $s > \tilde{s}$, using induction. Let $m(t)<0$ for all $t \in \{k_s+1,\ldots,k-1\}$, for some $k \in \{k_s+2,\ldots,k_{s+1}-1\}$. Repeated application of \eqref{eq:trust-update-subsequence-delta} gives
\begin{equation}
\label{eq:trust-update-subsequence-delta-general}
m(k) -m = \frac{k_s}{k} (m(k_s)-m) + \frac{1}{k} \sum_{t=k_s}^{k-1} (u_0(t)-m) + \frac{1}{k} \sum_{t=k_s}^{k-1} \triangle u(t)
\end{equation}
Along similar lines as \eqref{eq:delta-u-lipschitz}, one can show that
$|\triangle u(t)| \leq - \frac{m}{3}$ for all $t \in \{k_s, \ldots, k-1\}$. 
Using this, along with \eqref{eq:eps1} and \eqref{eq:eps3} ($\tau=k_s-1$, $r=k-k_s-1$), in \eqref{eq:trust-update-subsequence-delta-general} gives
\begin{equation*}
|m(k) -m| \leq - \frac{k_s}{k}\frac{m}{3} - \frac{(k+k_s-2)}{k}\frac{m}{6} - \frac{(k-k_s)}{k}\frac{m}{3} < - \frac{2m}{3}
\end{equation*} 
which implies that $m(k) < \frac{m}{3} \leq 0$. 


\end{document}